\documentclass[12pt,preprint]{aastex}
\def\lsim{\raise0.3ex\hbox{$<$}\kern-0.75em{\lower0.65ex\hbox{$\sim$}}}
\def\gsim{\raise0.3ex\hbox{$>$}\kern-0.75em{\lower0.65ex\hbox{$\sim$}}}

\begin{document}

\title{Cluster Formation in Protostellar Outflow-Driven Turbulence}
\author{Zhi-Yun Li$\!$\altaffilmark{1} \& Fumitaka 
Nakamura$\!$\altaffilmark{2}}
\altaffiltext{1}{Department of Astronomy, University of Virginia, 
Charlottesville, VA 22903; zl4h@virginia.edu}
\altaffiltext{2}{Faculty of Education and Human Sciences, Niigata 
University, 8050 Ikarashi-2, Niigata 950-2181, Japan; 
fnakamur@ed.niigata-u.ac.jp}

\begin{abstract}
Most, perhaps all, stars go through a phase of vigorous outflow during 
formation. We examine, through 3D MHD simulation, the effects of 
protostellar outflows on cluster formation. We find that the initial 
turbulence in the cluster-forming region is quickly replaced by motions 
generated by outflows. The protostellar outflow-driven turbulence 
(``protostellar turbulence'' for short) can keep the region close to 
a virial equilibrium long after the initial turbulence has decayed
away. We argue that there exist two types of turbulence in star-forming
clouds: a primordial (or ``interstellar'') turbulence and a protostellar 
turbulence, with the former transformed into the latter mostly in 
embedded clusters such as NGC 1333. Since the majority of stars 
are thought to form in clusters, an implication is that the stellar 
initial mass function is determined to a large extent by the 
stars themselves, through outflows which individually limit the mass 
accretion onto forming stars and collectively shape the environments 
(density structure and velocity field) in which most cluster members
form. We speculate that massive cluster-forming clumps supported by 
protostellar turbulence gradually evolve towards a highly centrally 
condensed ``pivotal'' state, culminating in rapid formation of massive 
stars in the densest part through accretion. 
\end{abstract}

\section{Introduction}

The origin of the stellar initial mass function (IMF) is a fundamental
unsolved problem in star formation. An important clue comes from 
millimeter 
and submillimeter observations of the nearby cluster-forming region,  
rho Oph cloud core, which uncovered prestellar condensations with a mass 
spectrum 
that resembles the Salpeter IMF (Motte et al. 1998; Johnstone et 
al. 2000; Stanke et al. 2005; see Testi \& Sargent 1998 for a similar 
result for the Serpens core). 
The implication is that the IMF may be determined to a large extent 
by the core mass distribution from cloud fragmentation. An 
attractive scenario is turbulent fragmentation. Padoan \& Nordlund 
(2002) showed analytically 
that, under plausible assumptions, an IMF-like core mass distribution 
can arise naturally from random turbulent compression. This result is 
corroborated by numerical simulations that include either driven (Li et 
al. 2004) or decaying turbulence (Tilley \& Pudritz 2005; see, however, 
Ballesteros-Paredes et al.  2005). In this picture, the IMF is 
determined largely by the properties of the turbulence. Since  
most stars are thought to form in clusters (Lada 
\& Lada 2003), the origin of the majority of stars boils down 
to the origin of turbulence in cluster-forming regions. 

It is established that supersonic turbulence decays quickly, on a 
time scale comparable to the turbulence crossing time on the dominant 
energy-carrying scale, with or without a strong magnetic field 
(Mac Low et al. 1998; 
Stone, Ostriker \& Gammie 1998; Padoan \& Nordlund 1999). In a turbulent 
cluster-forming region, if the dissipated energy is not replenished 
quickly, the cloud would be in a state of low level of turbulence and/or 
global free-fall collapse, neither of which is commonly observed 
(Evans 1999; Garay 2005). The 
turbulence must therefore be replenished somehow. The most likely 
mechanism is through (proto)stellar outflows, which are observed in 
abundance in nearby embedded clusters such as NGC 1333 (Knee \& Sandell 
2000; Bally et al. 1996). 

The idea of outflow-regulated star formation goes back to Norman \& 
Silk (1980). They envisioned that star-forming clouds are constantly 
stirred up by the winds of optically revealed T Tauri stars. This 
scenario was strengthened by the discovery of molecular outflows, 
which point to even more powerful winds from the stellar vicinity 
during the protostellar phase of star formation (Lada 1985; Bontemps 
et al. 1996). 
The protostellar outflows were incorporated into a general theory 
of photoionization-regulated star formation in magnetic clouds by 
McKee (1989). Their effects on cluster formation were examined 
by Matzner \& McKee (2000) using a parametrized rate of turbulence 
decay and a time dependent form of the virial 
theorem that treats the cloud as a whole. The cloud internal 
dynamics, such as mass distribution and velocity field, remain to 
be quantified. 

In this letter, we present a 3D MHD simulation of cluster formation 
in turbulent, magnetized clouds including protostellar outflows. In
agreement with previous work, we find that the initially imposed  
turbulence decays away quickly. We demonstrate that it is replaced 
by outflow-generated motions, which keep the cluster-forming region 
in an approximate virial equilibrium and allow for gradual star 
formation long after the decay of initial turbulence (\S~2). We argue 
in \S~3 that it is the protostellar outflow-generated turbulence 
(``protostellar turbulence'' hereafter), as opposed to the primordial 
(or ``interstellar'') turbulence that prevails in regions of molecular 
clouds that are little affected by local star formation activities, 
that is directly relevant to the formation of the majority of stars, 
including massive stars. 

\section{Protostellar Turbulence: An Example}

We illustrate the outflow-generated protostellar turbulence associated 
with cluster formation using a specific numerical example. Extensive 
parameter exploration is described in Nakamura \& Li (2006). 

We consider a centrally condensed cloud of density profile 
$ 
\rho (r) = \rho_0 [1+(6 r/L)^2]^{-1}
$ 
inside a sphere of radius $L/2$ (where $L$ is the size of our cubic 
computational box) and constant density outside. We pick a box size 
9 times the Jeans length 
$
L_J= (\pi c_s^2/G \rho_0)^{1/2}, 
$
where $\rho_0$ is the central density and $c_s=0.27~(T/20~{\rm K})^{1/2}$~km/s
the isothermal sound speed. Scaling $L_J$ by $0.17$~pc, we have a 
cloud size $L= 1.5 (L_J/0.17{\rm pc})$~pc, central number density 
$n_{H_2,0}=2.69 \times 10^4 (T/20~{\rm K}) (0.17{\rm pc}/ L_J)^2 
{\rm cm}^{-3} $, and mass $M=9.39\times 10^2 
(T/20~{\rm K}) (L_J/0.17~{\rm pc}) M_\odot$, all 
typical of nearby cluster-forming 
clumps (e.g., Ridge et al. 2003). We use as time unit the gravitational 
collapse time 
$
t_g= L_J/c_s = 6.12 \times 10^5 (L_J/0.17~{\rm pc})
(20~{\rm K}/T)^{1/2} {\rm yr}.
$
It is 3.27 times the free fall time at the cloud center and about $25\%$ 
longer than that at the average cloud density. Periodic conditions are 
imposed at the cloud boundaries. 

At the beginning of simulation, we impose on the cloud a uniform field 
along the $x$ axis, a moderate rotation along the $z$ axis, and a 
strong turbulence. The field strength is chosen to be 
$
B_0=7.48 \times 10^{-5} (T/20~{\rm K})(0.17~{\rm pc}/L_J)~{\rm Gauss}. 
$
The cloud is magnetically supercritical as a whole, with a flux-to-mass 
ratio 0.19 times the critical value $2\pi G^{1/2}$ in the central flux 
tube. The cloud rotation is specified by  
$
V_{\rm rot}=3 c_s (6 \varpi/L)[1+(6 \varpi/L)^2]^{-1}
$
inside the sphere of radius $L/2$ (where $\varpi$ is the distance from 
the rotation axis) and zero outside. The maximum rotation speed is 1.5 
times the sound speed. The cloud is stirred with a turbulent velocity 
field of power spectrum $v_k^2 \propto k^{-4}$ and rms Mach 
number $10$. Following the standard practice (e.g., Ostriker 
et al. 1999), random realization of the power spectrum is done in 
Fourier space. The turbulence is allowed to decay freely, except for
feedback from protostellar outflows. 

The evolution of the turbulent, rotating, magnetized cloud is followed 
using a 3D MHD code based on an upwind TVD scheme (see Nakamura \& Li 
2006). Our simulation has a relatively modest resolution of $128^3$. 
When the density in a cell 
crosses the threshold $\rho_{\rm th}= 100 \rho_0$, we define around
it a ``core'' that includes all cells in direct contact with the cell, 
either through a surface, a line or a point. The core 
is a cubic region having 3 cells on each side. We extract $20\%$ of
mass and momentum from the core, and put them in a Lagrangian 
particle at the core center. The particle represents a formed ``star'', 
whose subsequent motion in the cluster potential is followed 
numerically. The mass remaining in the core is assumed to be driven 
radially away in an outflow, with a momentum proportional to 
the stellar mass $M_*$. The proportionality constant $P_*$ 
(outflow momentum per unit stellar mass) is uncertain. Nakamura
\& Li (2006) estimated a plausible range of $10 - 100$~km/s. 
We adopt $P_*=50$~km/s, which is close to the value 40~km/s 
adopted by Matzner \& McKee (2000) in their semi-analytic model 
of cluster formation. 
Our recipe for mass extraction and protostellar outflow, while 
schematic, enables us to follow the cloud evolution well beyond 
the run-away collapse of the first density peak without violating 
the Jeans criterion (Truelove et al. 1997).

In Fig.~1, we plot the total kinetic energy of the gas $E_k$ as a 
function of time (in unit of gravitational collapse time $t_g$). In 
agreement with 
previous calculations, the turbulent energy decays quickly, to $\sim 
1/5$ of its initial value in as little as $0.1~t_g$. The decrease 
slows down at later times, partly because the decay time lengthens 
as the gas moves more slowly, and partly because of cloud contraction, 
which feeds gravitational binding energy into kinetic energy. By 
$0.4~t_g$, the first star has formed. It injects energy (and 
momentum) into the surrounding gas through an 
outflow, which shows up as the first spike on the kinetic energy 
curve in Fig.~1. Each of the subsequently formed stars also produces 
a spike. Together they reverse the decline of kinetic energy. The 
rms flow speed of the cloud reaches $\sim 1.5 (T/20~{\rm K})^{1/2}$~km/s 
toward the end of the simulation. The kinetic energy is to be compared 
with the gravitational energy $E_g$ of the system (also plotted in 
Fig.~1), which increases gradually with time as the cloud contracts. 
After the onset of star formation the kinetic energy remains 
comparable to the potential energy, indicating that the cloud is 
close to a virial equilibrium.

The quasi-equilibrium state is achieved through a realistic level 
of star formation. At the time $1.0~t_g$, there are 38 stars formed, 
with an efficiency (defined as the fraction of cloud mass turned 
into stars) of $2.1~\%$. The average mass of these stars is 
$\sim 0.5~M_\odot (T/20~{\rm K}) (L_J/0.17{\rm pc})$, which is 
typical of low-mass stars. The efficiency increases to $6.5~\%$ 
at $1.5~t_g$, and to $13.4~\%$ at $2.0~t_g$. The latter values are 
typical of the efficiencies inferred for nearby embedded clusters 
(Lada \& Lada 2003). Note that star formation can be rather bursty, 
particularly at late times, when stars are formed at a relatively 
high rate. 
  
The cloud morphology is transformed by active star formation. In Fig.~2, 
we show maps of column density at two representative times: one before 
star formation at $0.34~t_g$, and the other at $1.48~t_g$, when 102 
stars have formed. Before star formation the cloud appears relatively 
diffuse, with substructures created mostly by the (``interstellar'') 
turbulent velocity field imposed at the beginning of cloud evolution 
and, to a lesser extent, by moderate gravitational contraction. As time 
progresses, the initial turbulence dissipates, leading to more and more 
dense gas accumulating near the bottom of the gravitational potential 
well. The bulk of the dense gas is prevented from collapsing promptly 
into stars by protostellar outflows. While waiting to collapse, it 
is stirred up repeatedly, and often shredded into pieces, some of 
which coalesce into larger fragments. The abundance of dense, highly 
fragmented gas in the lower panels attests to the churning of 
cluster-forming gas by protostellar outflows. The churning results 
in a ``protostellar turbulence'' that appears distinct from the 
``interstellar turbulence'' found before significant star formation 
or outside cluster-forming regions. The difference is quantified 
in Nakamura \& Li (2006).

\section{Implications on IMF and Massive Star Formation}

The protostellar turbulence found in our simulation may be identified 
with the nonthermal gas motions observed in embedded clusters such as 
NGC 1333, where $\sim150$ stars have already formed (Lada et al. 1996). 
Molecular line (e.g., Walsh et al. 2005) and dust continuum observations 
of the region (Sandell \& Knee 2001) have revealed an abundance of dense 
fragmented gas, which broadly resembles the distribution shown in panel 
(f) of Fig.~2. There is also plenty of evidence for outflow activities, 
both current and past. Knee \& Sandell (2000) mapped $\sim10$ overlapping 
CO molecular outflows in the central part of the region. They suggested 
that the dense star-forming gas is fragmented by outflows, as seen in 
our simulation. Bally et al. (1996) identified more than one dozen 
HH flows. These collimated jets, to be simulated in near future, can 
propagate far from the YSOs that drive them and deposit part of their 
energy and momentum outside the dense ridge where active star formation 
is taking place. Quillen et al. (2005) uncovered $\sim 20$ CO shells 
in $^{13}$CO data cube, which they 
interpret as ``fossil'' outflows from earlier star formation. 
In addition, there are filaments emanating from the dense ridge into the 
lower density envelope (A. Goodman, priv. comm.), similar to those shown 
in panel (f) of Fig.~2. They could be associated with the blowouts 
of outflow-driven shells, akin to Galactic HI chimneys. Taken together, 
these observations leave little doubt that the star formation in 
NGC 1333 is strongly affected by outflows. 

In cluster-forming regions like NGC1333, the initial 
``interstellar'' turbulence dissipates quickly, and is replaced 
by protostellar turbulence. The initial turbulence sets the 
conditions for the formation of the first stars in a cluster. The 
majority of the cluster members are produced, however, after the 
initial turbulence has already decayed away, and replaced by 
protostellar turbulence (see Fig.~1). Therefore, it is the 
protostellar, rather than interstellar, turbulence that is more 
directly relevant to cluster formation. Since most stars are 
thought to form in clusters (Lada \& Lada 2003), we conclude 
that the majority of all stars form in environments shaped by 
protostellar turbulence. In this picture, protostellar outflows 
play a dual role in determining stellar masses: individually, they 
limit the mass accretion onto their driving protostars (Matzner 
\& McKee 2000; Shu et al. 2004). Collectively, 
they quickly replace the interstellar turbulence in controlling 
the velocity field and mass distribution of the cluster-forming 
gas. In our view, the protostellar turbulence holds the key to 
understanding of the origin of the masses of the majority of 
stars, including perhaps massive stars. 
  
Massive stars are observed to form in clusters. The current debate
centers on the scenarios of competitive accretion and stellar
merger (Bonnell et al. 1998) and turbulent core accretion (McKee 
\& Tan 2002). 
Our calculations demonstrated that, in the presence of protostellar
outflows of reasonable strength, the gas in a cluster forming 
region can be kept close a virial equilibrium (see also Matzner 
\& McKee 2000). The virialized gas would be difficult to capture 
efficiently by the stellar seeds envisioned in the competitive 
accretion scenario, as emphasized by Krumhotz et al. (2005), 
although higher resolution and refined treatment of outflows 
are needed to draw a firmer conclusion. 

McKee \& Tan (2002) envisioned the formation of individual massive 
stars (of say $30 M_\odot$) in discrete massive cores (of say $60 
M_\odot$). The cores are assumed to be part of a mass spectrum
determined by a turbulence of unspecified origin. If the formation 
of massive stars in a cluster is preceded by a period of vigorous 
low-mass star formation, it is likely that the outflows from the
low mass stars would quickly transform the turbulence in the cluster 
into a protostellar turbulence. Some support for this assertion 
comes from observations of infrared dark clouds, which are thought 
to be progenitors of massive stars and clusters (Menten 2005). The 
majority do not show clear evidence of free-fall collapse (Garay 
2005; see however Wu \& Evans 2003 and Peretto et al. 2005 for 
examples of rapid infall), 
indicating quick replenishment of decayed turbulence, most 
likely by outflows. In some sources, outflow activities associated 
with low mass star formation are clearly detected (for recent 
examples, see Rathborne et al. 2005 and Beuther et al. 2005). 
If this is the case in general, protostellar turbulence may hold 
the key to understanding the origin of massive stars as well. 

We offer a speculative scenario of massive star formation in 
protostellar turbulence that has  
conceptual parallels to the standard picture of isolated low-mass 
star formation (Shu et al. 1987; Mouschovias \& 
Ciolek 1999), where a magnetically supported clump gradually 
evolves towards a ``pivotal'' state of power-law density 
distribution through ambipolar diffusion. The ``pivotal'' state 
separates the gradual phase of core formation from the more 
dynamic phase of protostellar collapse and mass accretion (Li 
\& Shu 1996). We postulate that if a cluster-forming clump 
is massive and tightly bound enough to withstand the disruption 
by outflows from low and intermediate mass stars, it may gradually 
evolve towards a similar ``pivotal'' state. The protostellar 
turbulence may play the role of clump-supporting magnetic field 
envisioned in the standard picture. The gravitational settling 
of clump mass towards the bottom of potential well between 
outflows may play the role of ambipolar diffusion in driving 
the (clumpy) clump-wide density distribution towards a 
power-law -- the initial (t=0) condition for massive star 
formation, by accretion, from inside out. 

In this picture, there is no need for a pre-existing massive core 
to form a massive star, although one can plausibly identify the
highest density peak at the bottom of the cluster potential well
as a McKee-Tan core, once the region has become observationally 
distinguishable from the background. By this time, the (small) 
core may already be well on its way to dynamical collapse (and 
thus short-lived), accelerated perhaps by outflow trapping at 
high densities which 
effectively shuts off turbulence replenishment. In this 
interpretation, the core is automatically created at a privileged 
location - near the clump's (gravitational) center. The formation 
of massive stars may then be regarded as the culminating event of 
the (dissipative) gas evolution in massive cluster-forming clumps, 
perhaps not 
dissimilar to the formation of massive black holes at the centers 
of galaxies. This picture is consistent with the observations 
that ultracompact HII regions are usually found at the peaks 
of massive dust cores (Garay 2005), and hot molecular cores are 
often surrounded by more massive clumps of approximately 
power-law density distribution (Cesaroni 2005). Detailed 
calculations, now underway, will be used to test this scenario. 

\acknowledgments 
This work was supported by Grant-in-Aid for Scientific 
Research (No. 15740117) 
of Japan and NSF AST-0307368.

\begin{figure}
\plotone{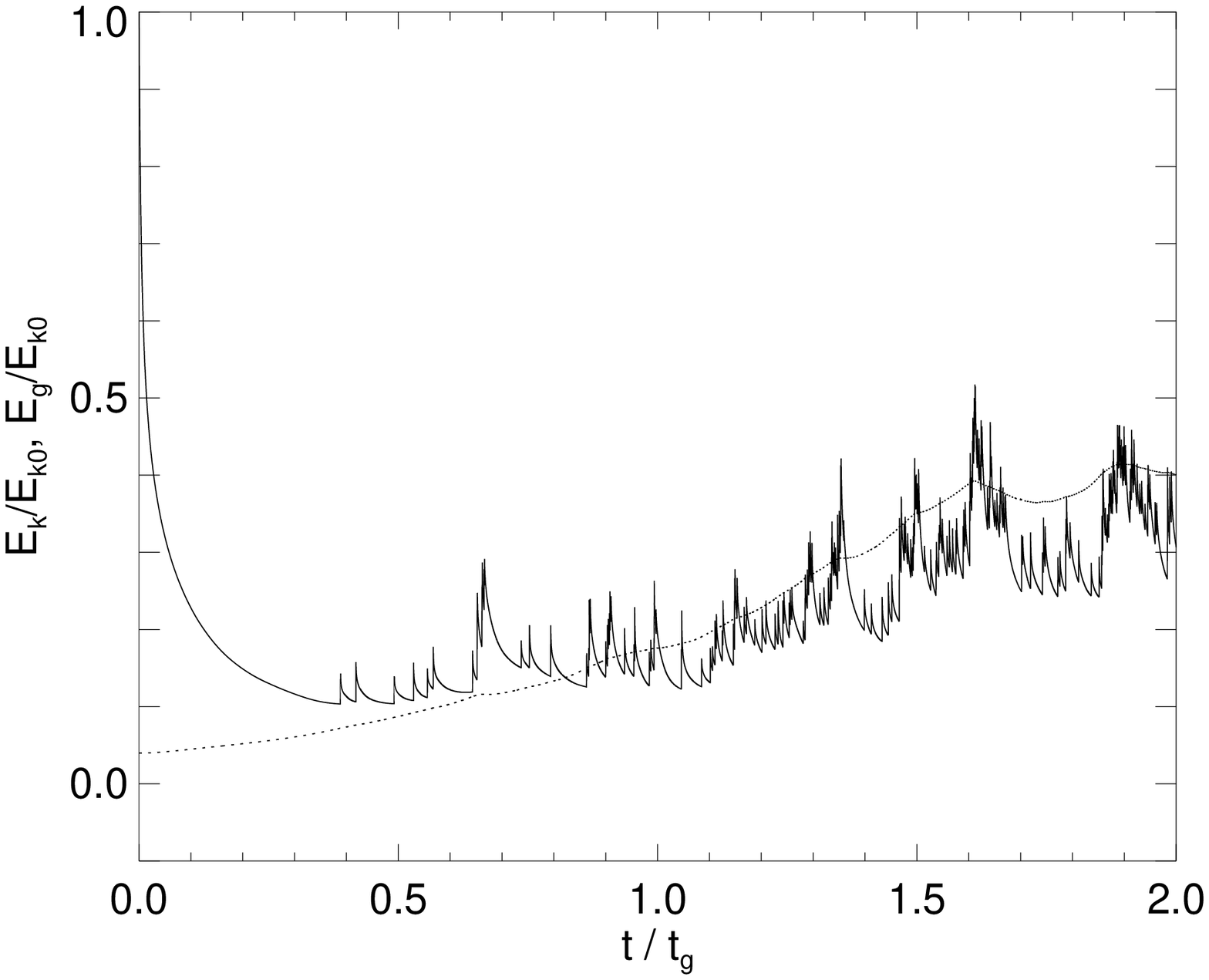}
\caption{Evolution of the total kinetic energy ($E_k$: {\it solid 
curve}) and gravitational energy ($E_g$: {\it dotted curve}). 
The energies are normalized to the initial kinetic energy 
$E_{k0}$ and the time to the gravitational collapse time $t_g$. }  
\end{figure}

\begin{figure}
\plotone{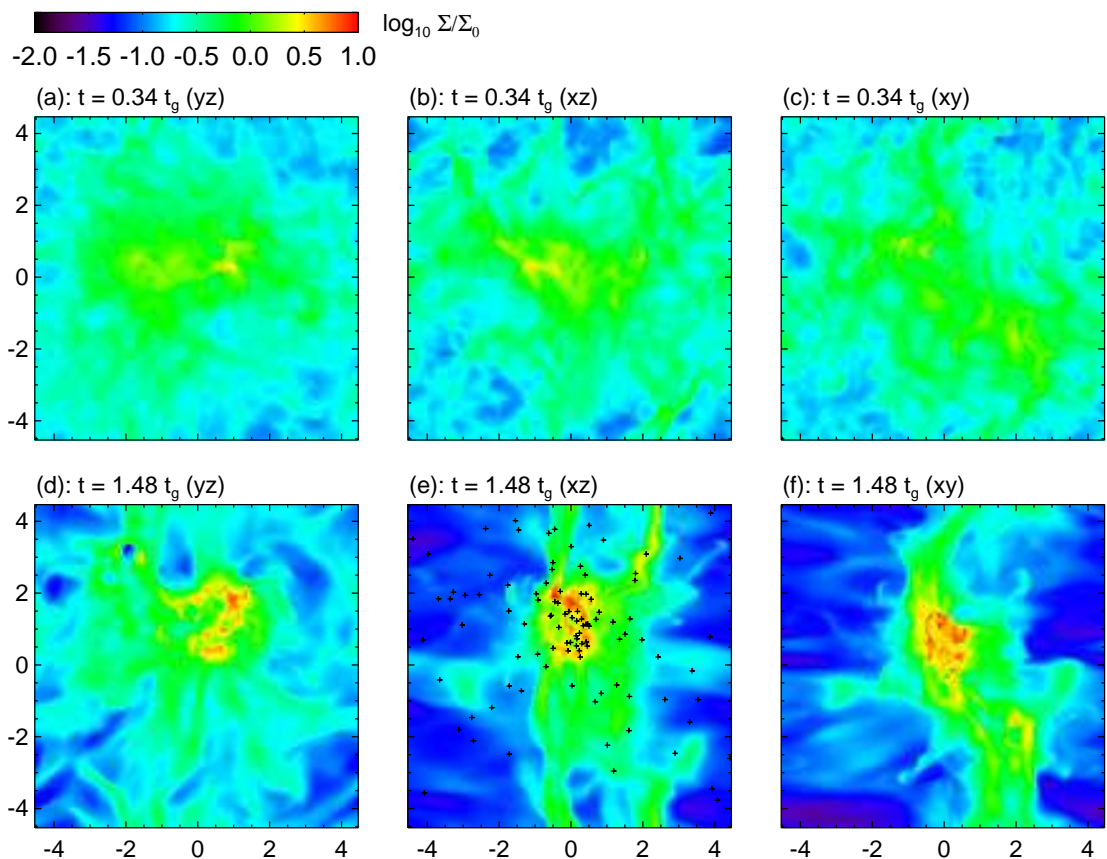}
\caption{Snapshots of column density distributions projected onto 
the $yz$ ({\it left panels}), $xz$ ({\it middle}), and $xy$ 
({\it right}) planes. The column density is normalized to the 
initial value through the cloud center $\Sigma_0$. The top panels 
are for the time $0.34~t_g$ 
before star formation, and low panels for $1.48~t_g$ after 102 
stars have formed. In panel (e), stars are marked by crosses.}
\end{figure}

\end{document}